# Orbital Multiferroicity in Pentalayer Rhombohedral Graphene

Tonghang Han[1]†, Zhengguang Lu[1]†, Giovanni Scuri[2,3], Jiho Sung[2,3], Jue Wang[2,3], Tianyi Han[1], Kenji Watanabe[4], Takashi Taniguchi[5], Liang Fu[1], Hongkun Park[2,3], Long Ju[1]*

[1]Department of Physics, Massachusetts Institute of Technology, Cambridge, MA, USA.

[2]Department of Chemistry and Chemical Biology, Harvard University, Cambridge, MA, USA.

[3]Department of Physics, Harvard University, Cambridge, MA, USA.

[4]Research Center for Electronic and Optical Materials, National Institute for Materials Science, 1-1 Namiki, Tsukuba 305-0044, Japan

[5]Research Center for Materials Nanoarchitectonics, National Institute for Materials Science, 1-1 Namiki, Tsukuba 305-0044, Japan

*Corresponding author. Email: longju@mit.edu †These authors contributed equally to this work.

**Ferroic orders describe spontaneous polarization of spin, charge, and lattice degrees of freedom in materials. Materials featuring multiple ferroic orders, known as multiferroics, play important roles in multi-functional electrical and magnetic device applications[1–4]. 2D materials with honeycomb lattices offer exciting opportunities to engineer unconventional multiferroicity, where the ferroic orders are driven purely by the orbital degrees of freedom but not electron spin. These include ferro-valleytricity corresponding to the electron valley[5] and ferro-orbital-magnetism[6] supported by quantum geometric effects. Such orbital multiferroics could offer strong valley-magnetic couplings and large responses to external fields—enabling device applications such as multiple-state memory elements, and electric control of valley and magnetic states. Here we report orbital multiferroicity in pentalayer rhombohedral graphene using low temperature magneto-transport measurements. We observed anomalous Hall signals $R_{xy}$ with an exceptionally large Hall angle ($\tan\Theta_H > 0.6$) and orbital magnetic hysteresis at hole doping. There are four such states with different valley polarizations and orbital magnetizations, forming a valley-magnetic quartet. By sweeping the gate electric field $E$ we observed a butterfly-shaped hysteresis of $R_{xy}$ connecting the quartet. This hysteresis indicates a ferro-valleytronic order that couples to the composite field $E \cdot B$, but not the individual fields. Tuning $E$ would switch each ferroic order independently, and achieve non-volatile switching of them together. Our observations demonstrate a new type of multiferroics and point to electrically tunable ultra-low power valleytronic and magnetic devices.**

Two-dimensional (2D) materials with honeycomb lattices feature electron valley as an internal degree-of-freedom that resembles spin. But due to its orbital nature, valley is easier to control through electric field, magnetic field and circular polarization of light[7–9]. The prospect of utilizing valley for information storage, transportation and processing has triggered intensive research in the field known as valleytronics[5]. In parallel, these materials host quantum geometric effects[6–10] that can be controlled by tuning the band structure, including the valley-dependent Hall effect and the orbital magnetism[6–10]. The latter could exhibit a much larger coupling to magnetic field than electron spin does[10,11]. These orbital-degrees-driven phenomena provide a fertile ground for novel ferroic orders, which warrants large responses to external fields and multi-functional device applications. For example, information could be stored in multiple states with different combinations of valley and magnetic characters instead of being limited to the binary states of spin. Furthermore, such multiple states can be manipulated conveniently by an electric field. Despite its great promise for applications and many theory proposals of realizing multiferroics in heterostructures of graphene and transition metal dichalcogenides[12–14], orbital multiferroics with independent switching of valley and magnetism have remained elusive.

Recently, spontaneous valley polarization and orbital magnetism induced by electron correlation effects have been observed in rhombohedral bilayer and trilayer graphene[15–19], as well as twisted graphene layers[20–24]. Despite the observed magnetic hysteresis, the valley and orbital magnetization are always locked. This is due to the requirement of a large gate electric field $E$ to generate flat bands for the Stoner instability[15–17], or the fixed band structure in a moiré superlattice[19–24]. The one-to-one correspondence between valley and magnetization prevents their independent control and realizing multiferroicity. In the twisted mono-bilayer graphene, flipping of orbital magnetization by charge doping in a fixed valley was observed[22]. The explanation relies on large edge-state-current contribution[25], while the magnetic moments in the bulk states are locked to valley. In a general material setting without moiré, it is crucial to understand the contributions of bulk states to ferro-valleytronic and ferro-orbital-magnetic orders without the complications of edge states. However, orbital multiferroicity in natural crystals has not been observed.

Here we explore orbital multiferroicity in pentalayer graphene with rhombohedral stacking. We purposely twist the angle between graphene and hBN layers away from zero degree to avoid moiré effects[15,26]. As shown in Figure 1a, orbitals located at the two highlighted sublattices dominate the electronic states near zero energy. Figure 1b shows the band structure and Berry curvature distribution in pentalayer graphene based on a tight-binding calculation[27]. At $E = 0$ V/nm, the two bands feature great flatness and a significantly larger density-of-states than those in thinner graphene (see Extended Data Fig. 3). At the same time, inversion symmetry of the system ensures a zero Berry curvature. When the bandgap is opened by $E$, states near the band edges acquire a non-zero Berry curvature[10,27–29]. The signs of the Berry curvature, the anomalous Hall resistance $R_{xy}$ and the orbital magnetization depend on the valley index and the sign of $E$. These results clearly show that valley and orbital magnetism are two separate order parameters in rhombohedral graphene. It is therefore possible to have a quartet of valley-polarized states, (K, +$M$), (K, -$M$), (K', +$M$) and (K', -$M$), each with the Fermi surface cuts through one of the four bands. The flat bands with tunable Berry curvature in pentalayer graphene provide a fertile ground for electron correlation and quantum geometry effects. The spin character of bands is not shown as it is irrelevant to the observations and discussions.

**Spontaneous ferro-orbital-magnetism**

Figure 2a shows $R_{xx}$ as a function of the electron density $n_e$ and $E$ in zero magnetic field. With hole doping in the flat band, a bubble-shaped region ('bubble' for simplicity) that is symmetric about $E = 0$ emerges. With an out-of-plane magnetic field $B = 1.8$ T, quantum oscillation features appear within the bubble as shown in Fig. 2b. The period of oscillation corresponds to two isospin flavors at the Fermi level (see Extended Data Fig. 2), indicating an isospin-symmetry-broken half-metal state. This bubble is absent from previous reports of rhombohedral graphene[15–19,30,31].

Figure 2c shows the $R_{xy}$ at $B = 0.5$ T. Outside the bubble, a small $R_{xy}$ signal prevails and corresponds to the normal Hall signal. Within the bubble, large anomalous Hall signals appear in a wings-shaped region ('wings'). The wings largely overlap with the region that shows quantum oscillations in Fig. 2b. Figure 2d shows $R_{xy}$ at several different $E$s at $n_e = -0.55*10^{12}$ cm$^{-2}$, as $B$ is scanned back and forth. No hysteresis is seen at $E = 10$ mV/nm, while clear hysteresis can be seen within the wings. The coercive field $\Delta B_c$ decreases while the jump in Hall resistance $\Delta R_{xy}$ increases, as $E$ is increased in amplitude. The values of $\Delta B_c$ and $\Delta R_{xy}$ as a function of $E$ are summarized in Fig. 2e. We note that the anomalous Hall angle is exceptionally large, with the largest tan$\Theta_H > 0.6$[32,33].

Combining the isospin-symmetry-breaking with the anomalous Hall signal, we conclude that the wings is valley-polarized. The magnetic hysteresis loop demonstrates a ferro-orbital-magnetism, where spin is not involved. Fig. 2d illustrates the states at large $B$ values. The Fermi surface is highlighted as a dashed circle, which cuts through the valence band in only one valley. Qualitatively, sweeping $B$ switches both the valley and magnetization at the same time. By switching the sign of $E$ at fixed $B$, the valley order is flipped while the magnetization remains since the magnetic field always chooses the valley that minimizes the magnetic energy. Microscopically, the close-to-linear relation between $\Delta R_{xy}$ and $|E|$ is due to the field-controlled Berry curvature distribution and can be reproduced by our tight-binding calculation

(see Extended Data Fig. 4). This demonstrates a new mechanism to achieve electrical tuning of orbital magnetism, while its counterpart for spin magnetism has been demonstrated[34]. We note the value of $\Delta B_c$ and $\Delta R_{xy}$ change in opposite ways when $E$ is changed, as the product $M\cdot B$ (where $M$ is the averaged orbital magnetic moment) needs to overcome an energy barrier to have the magnetization flipped. The value of $M$ is hard to be measured directly through transport measurements due to the small size of our device, but can be extracted indirectly as shown later.

The valley-polarized half-metal state in pentalayer graphene is induced by the electron-correlation effect in the flat band, in the form of a valley-exchange interaction[35–37]. Although other isospin-symmetry-broken states have been observed in Bernal bilayer graphene and ABC trilayer graphene (only at large $E$)[15–17], there was no sign of valley-polarized half-metal states. In addition, this state resides at around zero $E$, which leads to an important consequence as shown next.

**Ferro-valleytricity**

We further explore the valley and orbital magnetic orders by scanning the electric field $E$ at $B = 20$ mT, as shown in Fig. 3a&b. Surprisingly, the sign of $R_{xy}$ switches as the scanning direction switches if we focus on the wings. Fig. 3c shows a line-cut at $n_e = -0.55*10^{12}$ cm$^{-2}$, which reveals a butterfly-shaped hysteresis loop ('butterfly') within the wings. Repeating the measurements in Fig. 3c at –20 mT, a butterfly with the opposite winding direction appears as shown in Fig. 3e. We note that the small magnetic field is used to suppress the fluctuations of the whole magnetization (see Extended Data Fig. 8).

Based on earlier discussions, we illustrate the band structure and Fermi level corresponding to different states in Fig. 3c. When $E$ is scanned into the wings, the system selects the valley that minimizes the magnetic energy $-M\cdot B$. Scanning through zero $E$, the valley polarization persists at the cost of a higher magnetic energy (because $R_{xy}$ and orbital magnetization are flipped). Fig. 3d shows the valley polarization $V$ (defined as a +1 and –1 for K and K' valleys, respectively) extracted from Fig. 3c. It features a hysteresis loop and a new ferroic order we call ferro-valleytricity—the spontaneous polarization and switching of the valley order. Even at zero $E$, the valley polarization exists although the orbital magnetization is zero, in contrast to a non-zero magnetization at valley-polarized states in other graphene systems[15,19–23]. These observations show that the valley polarization and orbital magnetization are two different order parameters of the ground state, distinct from in the quarter-metal of trilayer graphene. Additionally, the butterfly hysteresis can exist in pentalayer graphene since the valley-polarized-half-metal resides at around zero $E$. While in rhombohedral trilayer graphene[15], the valley polarization is lost when $E$ is scanned to outside the quarter-metal state, and a butterfly was not observed.

We note the conjugate field of the valley polarization is $E \cdot B$. This can be seen from the winding directions of the valley-polarization in Fig. 3d&e: the two hysteresis loops under positive and negative $B$ wind in opposite ways, but can be unified if we plot $R_{xy}$ versus $E \cdot B$ instead of $E$ (see Extended Data Fig. 6). This is consistent with the relation between orbital magnetization and valley shown in Fig. 1b.

We provide a simple picture to understand both the ferro-orbital-magnetism and ferro-valleytricity observed in our device. The coupling between orbital magnetization and valley order can be described by a free energy $F = -\alpha V \cdot E \cdot B$ per hole, where $\alpha$ is a constant independent of $E$, $V$ and $B$. In the case of scanning $B$, $F$ can be viewed as $-(\alpha V \cdot E) \cdot B$, where $\alpha V \cdot E$ is effectively the averaged orbital magnetic moment and $B$ is the corresponding conjugate field. The averaged orbital magnetic moment $\alpha V \cdot E = g\mu_B$ ($\mu_B$ is the Bohr magneton and $g$ is the g-factor) will be extracted from a toy-model in the next section. In the case of scanning $E$, $F$ can be viewed as $-(\alpha V) \cdot (E \cdot B)$, where $\alpha V$ is effectively the valley moment and $(E \cdot B)$ is the corresponding conjugate field.

## $B$- and $T$-dependent ferro-valleytricity

With the ferro-valleytricity established, we further explore its response to the $B$ and elevated temperature $T$. Figure 4a&b show the evolution of butterfly at increased $B$, taken at $T = 300$ mK and $n_e = -0.55*10^{12}$ cm$^{-2}$ (see Extended Data Fig. 7 for other densities). At small $B$, the butterfly remains similar to that in Fig. 3c. As $|B|$ is increased, the range of $E$ showing hysteresis gradually shrinks and eventually disappears at ~ 0.6 T. Figure 4c shows line-cuts corresponding to the solid and dashed lines in Fig. 4a&b, where the butterfly gradually shrinks and evolves into an 'M' or 'W' shape. We define the critical electric field at which the butterfly terminates as $E_B$. At $B = -20$ mT, Fig. 4d shows $R_{xy}$ as $E$ is scanned at elevated $T$. Similar to its response to $B$, the butterfly gradually evolves into a 'W' shape as the temperature is increased to ~2 K. We define the critical electric field at which the butterfly terminates as $E_T$. Figure. 4e summarizes $E_B$ and $E_T$ as a function of $B$ and $T$, respectively.

A model to qualitatively describe the response to $B$ and $T$ is shown in Fig. 4f. The free energy of Ising-coupled orbital magnets has two local minima at $-M$ and $+M$ respectively, where $\pm M$ is the averaged orbital magnetic moment per hole. An energy barrier $\Delta$ prevents the flipping of the magnetization between $-M$ and $+M$. At $E_B$ and $E_T$, this barrier is overcome by a combination of the thermal energy $k_B T$ and the magnetic energy $M \cdot B$. As a result, the device switches from the local minimal energy state (K, -*M*) to the global minimal energy state (K', + *M*) and the butterfly is terminated. Further analysis of data in Fig. 4e allows us to extract the *g*-factor as shown in Fig. 4g (see Methods), which agrees with previous literatures[10,28]. Crucially, this averaged orbital magnetic moment per hole is ~10 times bigger than that of electron spin, which allows much stronger coupling with magnetic field.

**Orbital multiferroicity and switching**

The ferro-valleytronic order we observed enriches the family of ferroic orders in solids. Figure 5a compares ferro-valleytricity with conventional ferroic orders and conjugate fields. The valley order breaks both the inversion symmetry and the time-reversal symmetry, similar to the scenario of the ferro-toroidicity[38]. This can be seen from the (different form of) product of *E* and *B* as conjugate fields in both of cases. The coexistence of ferro-valleytricity and the orbital ferro-magnetism represents a new type of multiferroics where both orders ultimately originate from the orbital degree of electrons but not spin. These two orders couple strongly with each other, similar to that in type-II multiferroics[2]. Firstly, at fixed *B*, the conjugate field of the valley order (*E*•*B*) can continuously tune the orbital magnetic moment as shown in Fig. 3c and 4g, similar to the magneto-electric coupling in conventional type-II multiferroics. Secondly, the emergence of ferro-valleytricity and ferro-orbital-magnetism happen at a similar temperature of *T* = 2K, as can be seen in Fig. 4d and Extended Data Fig. 5. These co-incident critical temperatures are typical for type-II multiferroics, where one order spontaneously polarizes the other order through their coupling.

The gate electric field plays an important role in both ferroic orders as it contributes to the conjugate field of the valley order and controls the orbital magnetization. We can therefore use *E* to control the valley and orbital magnetization independently as shown in Fig. 5b. The valley-magnetic quartet states are connected by the butterfly in Fig. 3a, including (K, +*M*), (K, -*M*), (K', +*M*) and (K', -*M*). At small *B*, we can choose different routes along the butterfly to switch the valley repeatedly and keep the magnetization as shown in the blue-shaded part (Fig. 5b), or to keep the valley and switch the magnetization repeatedly as shown in the yellow-shaded part. In addition, the two ferro-magnetic states at the same *E*, for example (K, +*M*) and (K', -*M*) can be switched non-volatilely as shown in Fig. 5c.

The switching between the valley-magnetic quartet states requires only a change of the gate electric field, but not current flowing through graphene. This is very preferable to the current-induced switching of the magnetization from the power consumption point-of-view, which was the case in twisted bilayer graphene as well as conventional magnetic switches[20,21,39,40].

**Conclusion and outlook**

In summary, we demonstrated orbital multiferroicity in the flat bands of crystalline pentalayer graphene, which features co-existed but independent ferro-valleytricity and orbital-ferro-magnetism. It enriches the family of electronic phases of matter driven by the co-existing electron correlation and quantum geometry effects. Our experiment opens up new possibilities in harnessing the electron valley and orbital magnetism for multi-functional device applications in valleytronics and magnetics.

# Main Figures

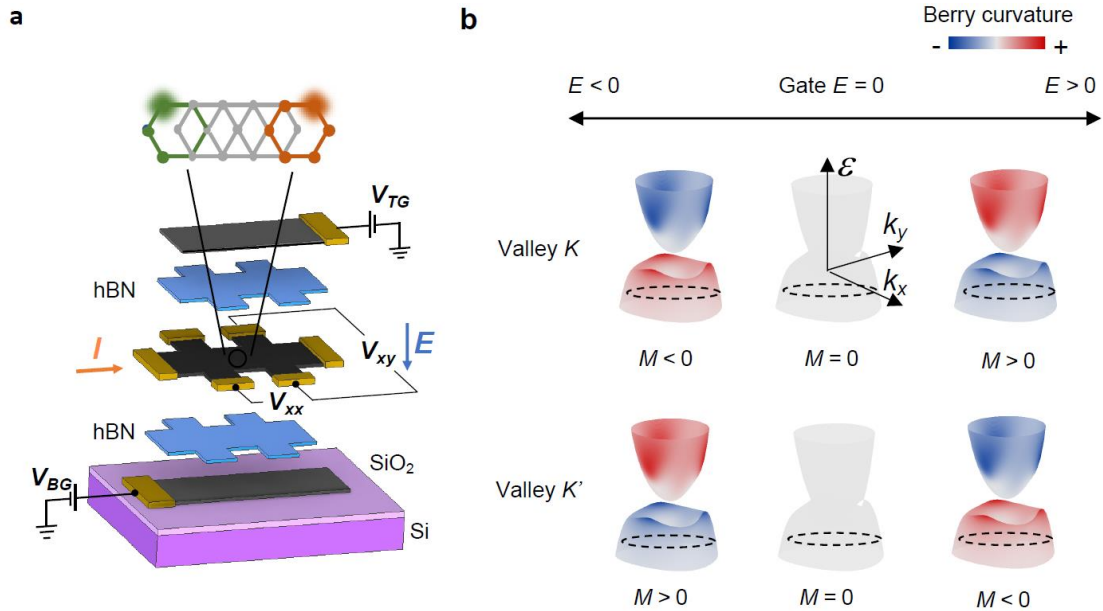

**Figure 1. Gate-induced Berry curvature and valley-magnetic quartet in pentalayer graphene with the rhombohedral stacking. a.** Schematic of the dual-gated Hall bar device, featuring the top view of the atomic structure of pentalayer rhombohedral graphene. The highlighted orbitals located at the bottom and top layers of pentalayer graphene dominate the wavefunctions of the low energy bands. **b.** Color-coded band structure in the K and K' valleys, showing the Berry curvature distribution at varied gate electric field $E$s. The valence band is flatter than the conduction band and more susceptible to electron correlation effects. The Berry curvature and orbital magnetic moment ($M$) of the hole states change signs as $E$ changes sign. Berry curvatures in the two valleys sum up to zero, to ensure the time-reversal symmetry. The four states at non-zero $E$ form a valley-magnetic quartet.

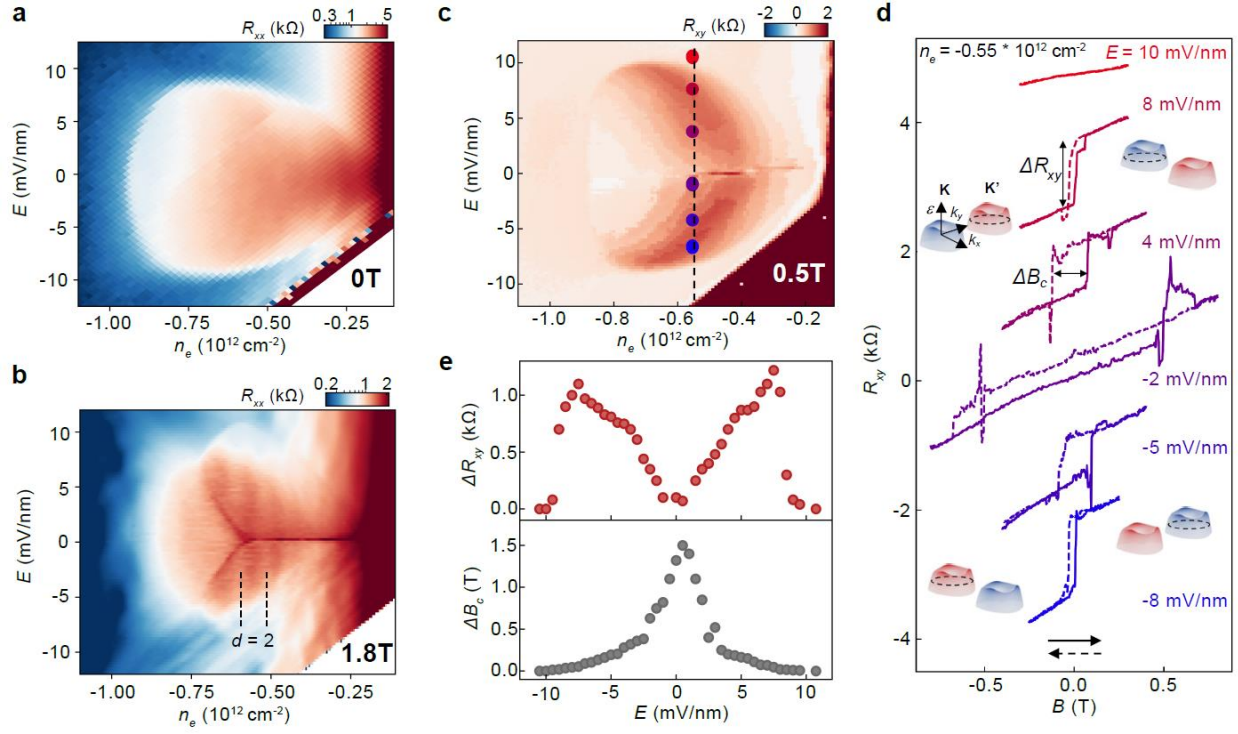

**Figure 2. Ferro-orbital-magnetism in a valley-polarized half-metal. a & b.** 2D color maps of the longitudinal resistance $R_{xx}$ at 0 T and 1.8 T out-of-plane magnetic fields. A bubble-shaped region appears when the flat valence band is doped by holes. At 1.8 T, quantum oscillations appear in part of the bubble and the period in density corresponds to two isospin flavors (degeneracy $d = 2$) at the Fermi level. **c.** 2D color map of the Hall resistance $R_{xy}$ at $B = 0.5$ T in the same range as in **a** & **b**. Anomalous Hall signal appears in a wings-shaped region which largely overlaps with where quantum oscillations are seen in Fig. 1d, indicating orbital magnetism due to valley polarization in this half-metal region. **d.** $R_{xy}$ as the magnetic field $B$ is scanned, showing clear hysteresis loops within the wings. Curves correspond to dots with the same color in **c** at $n_e = -0.55*10^{12}$ cm$^{-2}$ and different $E$s, and are shifted vertically for clarity. Solid and dashed curves correspond to forward and backward scanning of the $B$ field, respectively. Schematics of the valence band alignment and Fermi surface are shown for representative states. The bands are color-coded by the value of Berry curvature, the same as in Fig. 1b. **e.** The anomalous Hall signal $\Delta R_{xy}$ and the coercive field $\Delta B_c$ extracted from **d** along the dashed line in **c**. $\Delta R_{xy}$ is the amplitude of the abrupt jump of the transverse resistance which is extracted by averaging the changes of $R_{xy}$ at the two vertical edges of the magnetic hysteresis loop. $\Delta R_{xy}$ increases while the $\Delta B_c$ decreases with the increased amplitude of $E$.

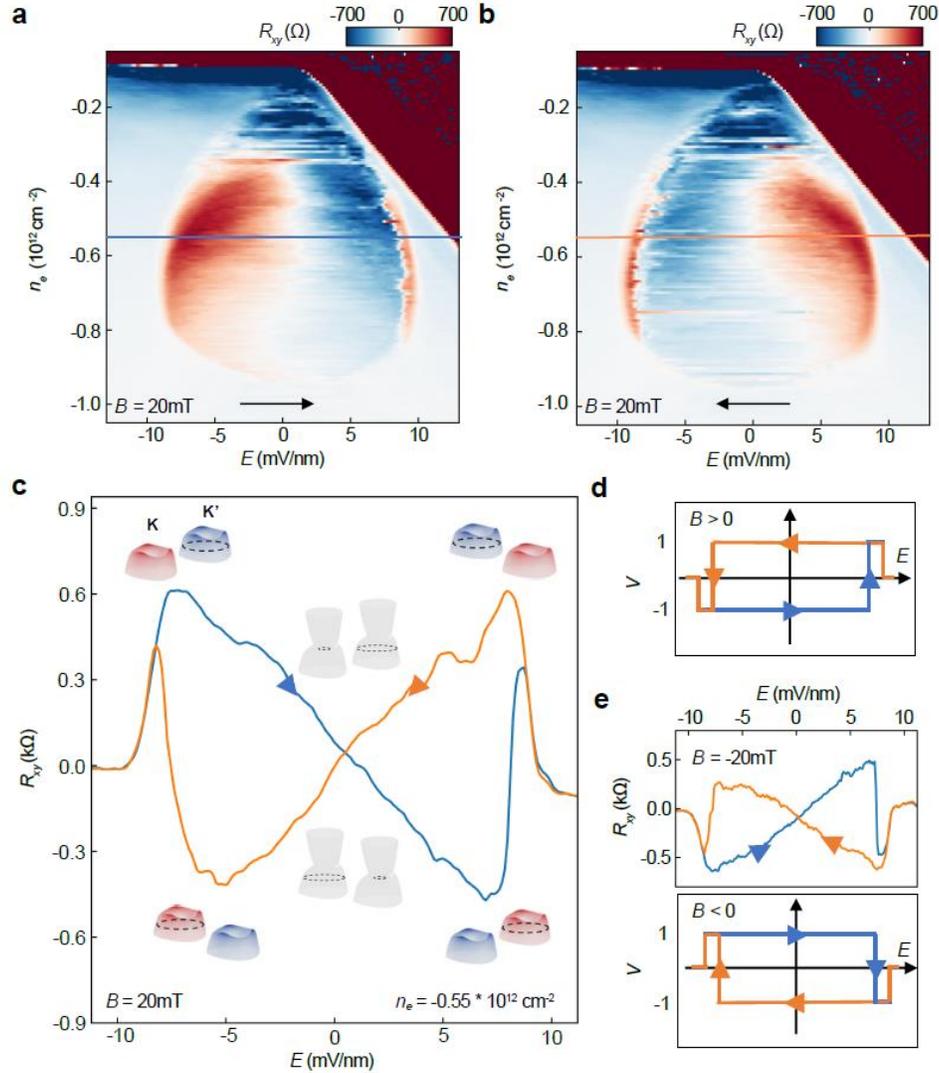

**Figure 3. Ferro-valleytricity. a** & **b.** 2D color maps of $R_{xy}$ corresponding to forward and backward scanning of $E$, at a small magnetic field $B = 20$ mT. **c.** $R_{xy}$ at $n_e = -0.55*10^{12}$ cm$^{-2}$, corresponding to lines shown in **a** & **b**. Clear hysteresis behavior is seen as a butterfly shape. This indicates the valley polarization persists as $E$ changes sign until close to the boundary and out of the wings. Schematics of the valence band alignment and Fermi surface are shown for representative states. The bands are color-coded by the value of Berry curvature, same as in Fig. 1b. The sharp spikes right before exiting the butterfly are due to the flipping of valley and $R_{xy}$ induced by the magnetic coupling -$M$•$B$, where $M$ is the averaged orbital magnetic moment. **d.** Plot of valley polarization $V$ corresponds to the $R_{xy}$ plot in **a**, featuring a hysteresis loop and a ferro-valleytronic order. **e.** Same plots as in **c** & **d** for a small negative magnetic field $B = -20$ mT. The direction of winding in $V$ is opposite of those in **d**, but they can be unified when plotting versus the conjugate field $E$•$B$. (See Extended Data Fig. 6).

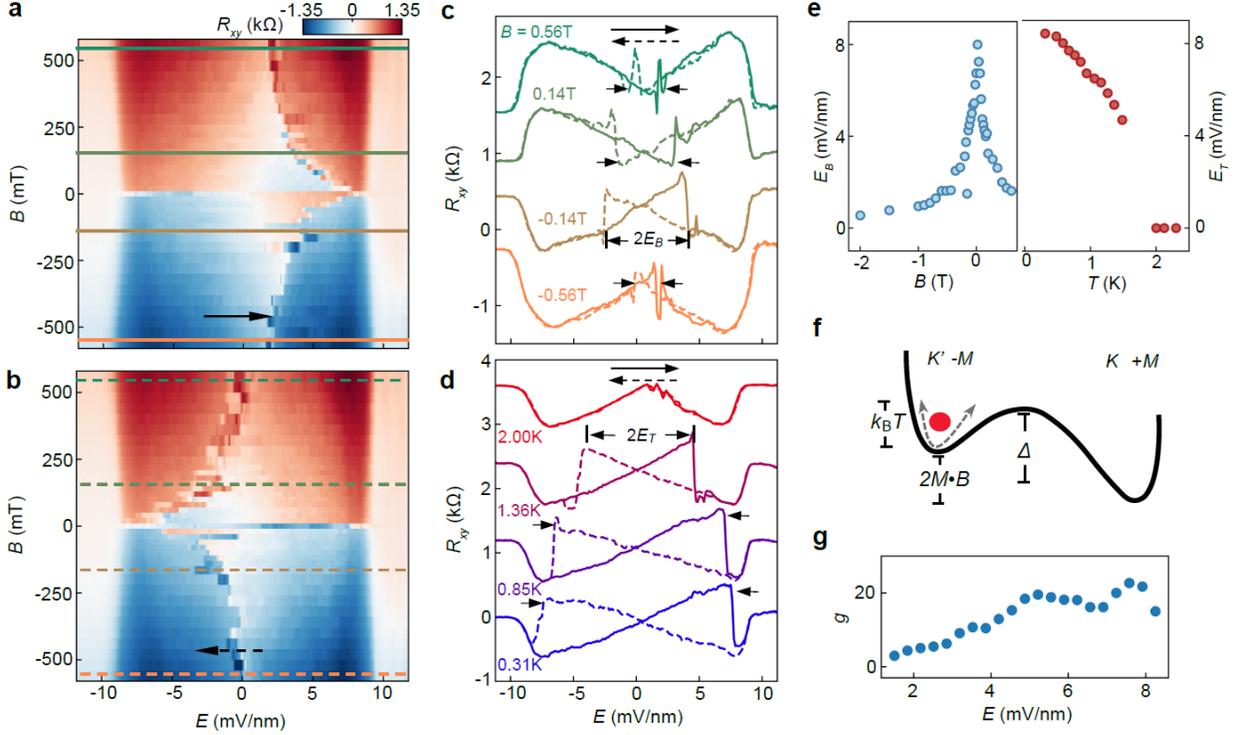

**Figure 4. Ferro-valleytricity controlled by magnetic field and temperature. a & b.** $R_{xy}$ as $E$ is scanned at different magnetic fields at $n_e = -0.55*10^{12}$ cm$^{-2}$. When $B$ is small, ferro-valleytricity dominates the device behavior. At above ~ 0.6 T, the $R_{xy}$ and $V$ are completely determined by the magnetic field while the butterfly disappears. **c.** Line-cuts corresponding to the solid and dashed lines in **a** & **b**, showing the butterfly gradually shrinks in $E$ and eventually evolves into an 'M' or 'W' shape. $E_B$ indicated by the arrows is the critical field at which the butterfly is terminated. **d.** $R_{xy}$ as $E$ is scanned at different temperatures at $n_e = -0.55*10^{12}$ cm$^{-2}$ and $B = -20$ mT. The butterfly and ferro-valleytricity gradually disappear as the temperature is increased, similar to the behaviors as when $B$ is increased. $E_T$ is the critical field at which the butterfly is terminated. **e.** Critical fields $E_B$ and $E_T$ from **c** & **d**, respectively. **f.** A schematic to explain the flipping of $R_{xy}$ at the critical fields $E_B$ and $E_T$. The flipping happens when the magnetic energy $M \cdot B$ and thermal energy $k_B T$ add up to overcome the energy barrier D. **g.** Effective g-factor of the averaged orbital magnetic moment as a function of $E$ extracted from **e**.

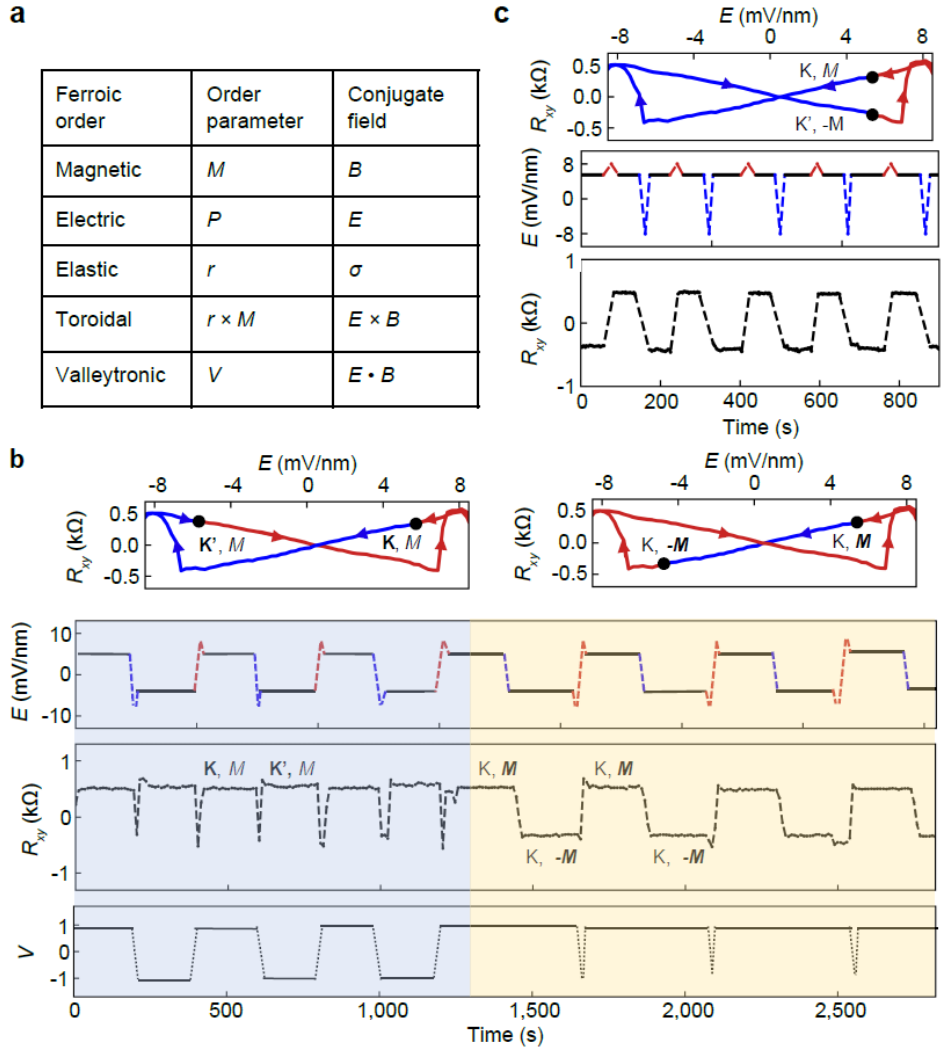

**Figure 5. Electrical control of the orbital multiferroic. a.** Ferroic orders and the corresponding conjugate fields. The ferro-valleytronic order breaks both inversion (due to $E$ field) and time-reversal symmetry (due to $B$ field), similar to the ferro-toroidal order does. The ferro-valleytronic order and ferro-orbital-magnetic order form an orbital multiferroic. **b.** Switching of the valley polarization $V$ (blue shaded region) and orbital magnetic moment $M$ (yellow shaded region) independently by sweeping the gate electric field $E$. The top panel shows the path of switching along the 'butterfly' between the two static states labeled by the two dots. The color on the paths matches that of the switching pulse and the arrows label the scanning direction. A small magnetic field $B = 20$ mT is applied here. **c.** Non-volatile switching of the $M$ and $V$ together using a gate electric field.

## Methods

### Device fabrication

The pentalayer graphene and hBN flakes were prepared by mechanical exfoliation onto $SiO_2$/Si substrates. The rhombohedral domains of pentalayer graphene were identified using near-field infrared microscopy[8], confirmed with Raman spectroscopy and isolated by cutting with a Bruker atomic force microscope (AFM)[41]. The van der Waals heterostructure was made following a dry transfer procedure. We picked up the top hBN and pentalayer graphene using polypropylene carbonate (PPC) film and landed it on a prepared bottom stack consisting of an hBN and graphite bottom gate. We intentionally misaligned the straight edges of the flakes to avoid the formation of the moiré superlattice. The device was then etched into a Hall bar structure using standard e-beam lithography (EBL) and reactive-ion etching (RIE). We deposited Cr/Au contact electrodes and a NiCr alloy top layer to form a dual-gate device.

### Transport measurement

The device was measured in a Bluefors LD250 dilution refrigerator with an electronic temperature of around 100 mK. Part of the data was taken in the National High Magnetic Field Laboratory, Tallahassee, FL. Stanford Research Systems SR830 lock-in amplifiers were used to measure the longitudinal and Hall resistance $R_{xx}$ and $R_{xy}$ with an AC voltage bias 200 uV at a frequency at 17.77 Hz. Keithley 2400 sourcemeters were used to apply top and bottom gate. Top-gate voltage ($V_t$) and bottom-gate voltage ($V_b$) are swept to adjust doping density $n_e = (C_t V_t + C_b V_b)/e$ and electric field $E = (V_t/d_t - V_b/d_b)/2$, where $C_t$ and $C_b$ are top- and bottom-gate capacitance per area calculated from the Landau fan diagram

### Phase diagram of pentalayer rhombohedral graphene

Extended Data Fig. 1a and 1b show the color plots of four-probe resistance $R_{xx}$ as a function of carrier density $n$ and displacement field $E$ for the hole doping side and electron doping side measured at $B = 0$ and a temperature of 100 mK. Colored dots label different phases including band insulator (BI), correlated insulator (CI), spin-polarized half metal (SPHM), isospin-polarized quarter metal (IPQM), unpolarized metal (UP) and valley-polarized half metal (VPHM). The large $D$ side of the phase diagram looks similar to that of the previous studies in bilayer and trilayer graphene[15,16]. At the intermediate and small $D$ side, the semimetal and correlated insulator states emerge and dominate the neighboring regions. The VPHM state with a waterdrop shape (labeled by a red star) is the focus of this paper.

    The degeneracy of the isospin-polarized metal is inferred from the spacing of the Landau levels[15]. Extended Data Fig. 2a and 2b show the $R_{xx}$ as a function of $n$ and $E$ measured at $B = 2T$ in the VPHM

state and the large density UP state. The Landau level spacings indicate the Fermi surface has 2-fold and 4-fold degeneracy in the VPHM and UP state respectively. The details of IPQM and SPHM states can be found in ref[42].

**Tight-binding calculation of band structure and Hall conductivity**

The single-particle band structure of the rhombohedral stacked pentalayer graphene is calculated from the 10-band continuum model. The Hamiltonian and the parameters are taken from previous literature[43]. The band structures and density of states (DOS) calculated for different layer number $N$ (2$N$-band model) are shown in Extended Data Fig. 3. The remote hopping processes lead to the trigonal warping of the band structure, resulting in deviation from $E \sim k^N$ at low energy. We note that among all layer numbers, rhombohedral pentalayer graphene has the flattest band and largest DOS around zero energy. This could account for the observation of new correlated phases absent in Bernal bilayer and rhombohedral trilayer graphene.

Extended Data Fig. 4a shows the tight-binding calculation of band structure for rhombohedral pentalayer graphene near the K point with an interlayer potential $2E_0 = 0$ meV (black) and 16 meV (grey). Inset shows an iso-energy contour at –5 meV and the three-fold rotational symmetry is from the trigonal warping effect. Due to the trigonal warping, the Dirac points arrange themselves along the three directions and at the center[29]. The black and white dots represent Dirac points with Berry phase $\pi$ and $-\pi$, which are also labeled by the black and white arrows in the band structure. They appear at different energy and correspondingly different densities.

We also calculated the intrinsic contribution to the anomalous Hall conductivity by integrating the Berry curvature of the occupied states $\sigma^{AHE} = \int_{occupied\ states} \Omega$. Extended Data Fig. 4b shows the calculated Hall conductivity with respect to the doping density $n_e$ and interlayer potential $E_0$ for a single valley and spin flavor. The hot spots at around -1.2 and 0.4*10$^{12}$ cm$^{-2}$ correspond to the four Dirac points at smaller $k$ and the three Dirac points at larger $k$ respectively. In our valley-polarized half-metal picture, one valley has holes while the other valley has zero net carriers ($n_e$=0). As a result, to fully characterize the system we need to add the contributions from the two valleys together. Extended Data Fig. 4c shows the same plot as **b** with the contribution from the other valley at zero density (opposite value to the dashed line in **b**). Extended Data Fig. 4d provides a microscopic picture for the linear anomalous Hall conductivity with $E$. At $E = 0$, Berry curvature is zero everywhere except for at the Dirac points, so $\sigma^{xy}$ is zero. As $E$ increases, Berry curvature hot spots emerge near the Dirac points and spread out. Consequently, there is more Berry curvature on states below $E_F$ and $\sigma^{xy}$ increases with $E$. Extended Data Fig. 4e shows linecuts in **c** from $n_e$= 0 to -0.5*10$^{12}$ cm$^{-2}$ for a single spin copy. At a small interlayer potential $E_0$, the Hall conductivity is roughly linear with $E_0$. Extended Data Fig. 4f shows measured $\sigma^{xy}$. The linear $E$ dependence of $\sigma^{xy}$ agrees with the calculation qualitatively. We note that the measurements on the positive $E$ side suffer from contact issue at low temperatures (can be seen from the $n$-$E$ color plot as well). The measured $\sigma^{xy}$ is much larger than the calculated Berry curvature contribution, which could arise from extrinsic contribution[44]. This phenomenon is also seen in other graphene based orbital magnet, for example the quarter-metal of rhombohedral trilayer graphene[15].

**Doping dependence of the orbital magnetism**

We show the electric field dependence of the anomalous Hall effect (orbital magnetism) in the main text and here we present the doping dependence. Extended Data Fig. 5a is the same as Fig. 2c and now we cut along the x-axis. Extended Data Fig. 5b and 5c are magnetic hysteresis measured at $E$ = 5 mV/nm and $n_e$ = -2.5 to -9.5*$10^{11}$ cm$^{-2}$ (corresponds to the dots with the same color in 5a) with temperatures of 0.3K and 2K. Clear magnetic hysteresis and anomalous Hall effect are seen within the waterdrop region and we get the largest signal at around $n_e$ = -5.5*$10^{11}$ cm$^{-2}$. We also note that the magnetic coercive field at 2K almost vanishes, aligned with the disappearance of butterfly at 2K.

**Control the valley order with $E·B$ field**

Here we demonstrate $E·B$ as the conjugate field of the valley order. Extended Data Fig. 6a and 6b plot show the valley polarization corresponding to the $R_{xy}$ plot in Fig. 3d & 3e with $B$>0 and $B$<0. The two hysteresis loops wind in opposite ways. However, if we plot the x-axis as the $E·B$ field, the two loops will collapse into one, which demonstrates that the conjugate field of the valley is $E·B$. Blue (yellow) lines labels scanning to positive (negative) $E$ direction. The arrows indicate the scanning direction.

**Random fluctuation of magnetization at $B$=0**

Extended Fig. 8 shows the fluctuation of $R_{xy}$ as a function of time at $B$ = 0 T. $R_{xy}$ flips sign frequently with time, indicating the fluctuation of the magnetization due to the finite size of the magnet. The stochastic switching enables the system to act as a probabilistic bit, which is essential in probabilistic and neuromorphic computing[45,46]. Crucially, a small external $B$ field (20 mT) is enough to stabilize the system to a controlled state without changing the ground state. As a result, our device is capable of shifting between two operation modes (stochastic and controlled) easily.

**$R_{xy}$ measurement under different conditions**

There are several measurement conditions including scanning doping density or electric field, and applying a small positive or negative $B$ field. In the main text Fig. 3a and 3b we show the plots of scanning $E$ with a positive magnetic field. Here in Extended Data Fig. 9 we discuss the results of other measurements (the applied magnetic field is ±20 mT) and all four measurements are consistent with our picture.

   i. **Scan $E$, negative $B$** (9a, 9b). Similar to the main text Fig. 3a and 3b, the negative magnetic field now selects the opposite valley compared to the former case. The behavior is the same as Fig.3a and 3b with an opposite sign of $R_{xy}$.
   ii. **Scan $n_e$, positive $B$** (9c, 9d). In contrast to the scanning $E$ case, there is no butterfly-type hysteresis when scanning $n$, since the orbital magnetism of a valley does not change with $n$. The system will choose to polarize to the valley with magnetization parallel to $B$, resulting in positive $R_{xy}$ in both positive and negative $E$ sides. The magnetization of the system is small around $E = 0$ and the small external $B$ field cannot effectively select a favored valley, resulting in fluctuations around $E = 0$. We note the fluctuations around $E = 0$ do not happen when scanning $D$ since the system already chooses the valley as the system enters the VPHM state and stays in the chosen valley.
   iii. **Scan $n_e$, negative $B$** (9e, 9f). Similar to the 9c and 9d but with the opposite external $B$ field. The small negative magnetic field will select the opposite valley compared to the 9c & 9d case.

**Hysteresis of the longitudinal resistance $R_{xx}$**

In the main figures we mainly show the $R_{xy}$ behaviors and relate the jump of $R_{xy}$ to the switching of the valley. Here we also present the $R_{xx}$ behaviors corresponding to the $R_{xy}$ and show the $R_{xx}$ jump simultaneously with $R_{xy}$.

Extended Data Fig. 10a-f show the n-E map with a small positive magnetic field. 10a and 10b are the same as Fig. 3a and 3b. 10c and 10d are the $R_{xx}$ map corresponding to 10a and 10b. The white arrows indicate the jumps at the boundary, which matches where $R_{xy}$ jumps. 10e and 10f are linecuts at $n_e$=-0.67*10$^{12}$cm$^{-2}$, indicated by the dashed line in 10c and 10d. The arrows label the scanning directions. The jumps in $R_{xx}$ happen right before the system get out of the VPHM state where $R_{xy}$ also jumps. Extended Data Fig. 10g-l plots the data corresponding to the measurement in Fig. 4a and 4b, in a similar way to 10a-f. It is clear that the $R_{xx}$ jumps at the same time as $R_{xy}$ jumps. The hysteretic behaviors in $R_{xx}$ indicate the valley-switching is a first-order process.

**Extracting the averaged magnetic moment through a toy model**

In Fig. 4g we phenomenologically extracted the averaged magnetic moment per hole from the magnetic field dependence and temperature dependence measurement of the $R_{xy}$-$E$ butterfly. Our picture is depicted in Fig. 4f where we assume the barrier Δ does not change with magnetic field and temperature. Our sweeping of the magnetic field in Fig. 4 a&b was done very slowly (each scan takes ~minutes). Therefore, in our toy model the system will switch to the valley with lower energy (global minimum) when the magnetic energy added with thermal energy is larger than the barrier, i.e. $M·B+k_BT > Δ$. Then we can estimate the averaged magnetic moment $M$ at different $E$ with the extracted critical electric field $E_C$ at different magnetic fields and temperatures in Fig. 4e. We have

$M(E_B)·B(E_B)+k_BT(0.3K) = Δ$

$M(E_T)·B(0.02T)+k_BT(E_T) = Δ$

Where $E_B$ and $E_T$ are the critical electric fields when changing magnetic field and temperature. So the averaged magnetic moment can be written as

$M(E_C) = (k_BT(E_T) - k_BT(0.3K)) / (B(E_B) - B(0.02T))$

We extrapolate the data in Fig. 4e and calculate the averaged magnetic moment $M$ as a function of electric field and get Fig. 4g. We note that the absence of data at 1.5-2 K in Fig. 4e is due to the difficulty of stabilizing the sample temperature in this range in our experimental setups. When extracting the *g*-factor, we used the value of $E_T$ by linear interpolation in this range.

**Relation between ferro-valleytricity, ferroelectricity and magneto-electric effect**

The coupling term $F = - αE*V*B = -(αE*V)*B = - E*(αV*B)$ can be viewed as an orbital magnetic moment coupling to a magnetic field, as well as an electric polarization coupling to an electric field. In the main text we discussed the orbital magnetism and here we elaborate on the electric polarization.

We note that the total electric polarization/dipole has two parts. Without considering the valley symmetry breaking, an electric dipole is induced by the gate electric field $E$. At gate electric field $E = 0$ mV/nm, in both K and K' valleys, both the conduction and valence band wave functions have 50% at the top layer and 50% at the bottom layer, preserving the inversion symmetry of the lattice structure; when a non-zero $E$ is applied, the inversion symmetry is broken so in both valleys the valence-band-wave-function is polarized in the layer with a lower potential energy while the conduction-band-wave-function is polarized in the opposite layer. From the band structure point of view, applying a gate electric field opens up the band gap proportional to $E$ and lowers the energy of the valence band, as shown in Extended Data Fig. 11a. The total energy lowering of the valence band and occupied states is equivalent to an electrostatic energy of $E*P_0$, where the electric dipole $P_0$ is induced by the applied $E$ and characterizes the number of electrons whose energy is affected by the gap opening. Although the electric potential energy includes a term that is $–E*P_0$, we note that this energy should not be counted in the flipping process from

K' to K: as we explained above, $P_0$ is unrelated to valley polarization, so **all the states on the free-energy curve in Fig. 4f have exactly the same potential energy $-E*P_0$ as they correspond to the same $E$ and $P_0$ is unrelated to valley polarization.** As $P_0$ is unrelated to valley polarization, vice versa, the ferro-valley order does not imply a ferro-electric order, since the system can have a big(small) Fermi surface in the K(K') valley even if the inversion symmetry is preserved, as shown by the states at the center of the butterfly in Fig. 3c of the main text.

When the valley symmetry is spontaneously broken in the presence of a magnetic field $B$, the bands in K and K' valley shift in energy due to the coupling between the orbital magnetization and $B$. This coupling results in the free energy $F = -\alpha E*V*B = -(\alpha E*V)*B$, where $\alpha E*V$ is the averaged orbital magnetic moment as we stated in the manuscript. This same energy can be as $F = -E*(\alpha V*B)$, where $(\alpha V*B)$ is an effective electric dipole $P'$. This additional electric dipole $P'$ is determined by both $V$ and $B$, and it is involved in the ferro-valleytricity behavior. From the band structure point of view, this electric dipole and the corresponding electrostatic energy can be seen from the shift of the occupied states in valence band in a magnetic field $B$, as shown in Extended Data Fig. 11b.

In the presence of $B$, $P'$ (at $E = 0$ mV/nm) and valley polarization $V$ should indeed co-exist. Here we provide a physical picture to understand this (Extended Data Fig. 11c). At a non-zero $B$, electron states form discrete Landau levels. Taking the zeroth Landau level as an example (other Landau levels have qualitatively similar sublattice and layer polarization, but quantitatively not as big as that in the zeroth Landau level), its electron wavefunctions in the K and K' valleys are located at the A and B sublattices respectively. In all rhombohedral graphene, A and B sublattices to describe the lowest-energy bands are located at the top and bottom layers. Therefore, valley polarization leads to the imbalance of population of the zeroth Landau level, which results in a layer polarization of electric charge distribution and electric dipole.

In summary, ferro-valley order is not related to ferroelectricity when $B = 0$. Only with the presence of a magnetic field, the ferro-valley order leads to ferroelectricity in a way that $P' = (\alpha V*B)$. To probe this ferroelectric order, a graphene sensing layer or layer-sensitive capacitance measurement is required[47–49]. We also note that when considering the free energy, we cannot consider the orbital magnetism and electric polarization at the same time, since they come from essentially the same term $F = -\alpha E*V*B$ and we shouldn't double count this energy.

The induced electrical polarization $P'$ and orbital magnetic moment $M$ can be viewed as a magneto-electric effect, where the free energy $F = -\alpha V*B*E = -E*P' = -B*M$. The valley polarization $V$ can be viewed as the coefficient of the magnetoelectric effect, which usually can be controlled through strain in conventional multiferroic or magneto-electrics, for example.

**Density and in-plane magnetic field scan in the correlated insulating state at $D = n = 0$**

We performed similar measurements as in ref[18]. Although the back-and-forth scans do not overlap perfectly (Extended Data Fig. 12), we did not see obvious hysteresis as reported in ref[18].

## Methods references

**Acknowledgments**

We acknowledge helpful discussions with F. Zhang, J. Checkelsky, L. Levitov, Z. Dong, W. He, P. Yu, Y. Ba, F. Wang, and L. Zhao. We acknowledge Y. Yao for his help in sample fabrication. L.J. acknowledges support from a Sloan Fellowship. Work by T.H. was supported by NSF grant no. DMR-2225925. The device fabrication of this work was supported by the STC Center for Integrated Quantum Materials, NSF grant no. DMR-1231319 and was carried out at the Harvard Center for Nanoscale Systems and MIT.Nano. Part of the device fabrication was supported by USD(R&E) under contract no. FA8702-15-D-0001. K.W. and T.T. acknowledge support from the JSPS KAKENHI (Grant Numbers 20H00354, 21H05233 and 23H02052) and World Premier International Research Center Initiative (WPI), MEXT, Japan. L.F. was supported by the STC Center for Integrated Quantum Materials (CIQM) under NSF award no. DMR-1231319. H.K. acknowledges support from NSF grant no. PHY-1506284 and AFOSR grant no. FA9550-21-1-0216. A portion of this work was performed at the National High Magnetic Field Laboratory, which is supported by National Science Foundation Cooperative Agreement No. DMR-2128556* and the State of Florida.

**Author Contributions**

TH.H., Z.L., G.S., J.S. and J.W. performed the DC magneto-transport measurement. TH.H. and TY.H. fabricated the devices. K.W. and T.T. grew hBN single crystals. TH.H. performed the tight-binding calculations. H.P. and L.F. contributed to the data analysis. L.J. supervised the project. All authors discussed the results and wrote the paper.

**Competing Interests** The authors declare no competing interests.

**Correspondence and requests for materials** should be addressed to Long Ju.

**Extended Data Figures**

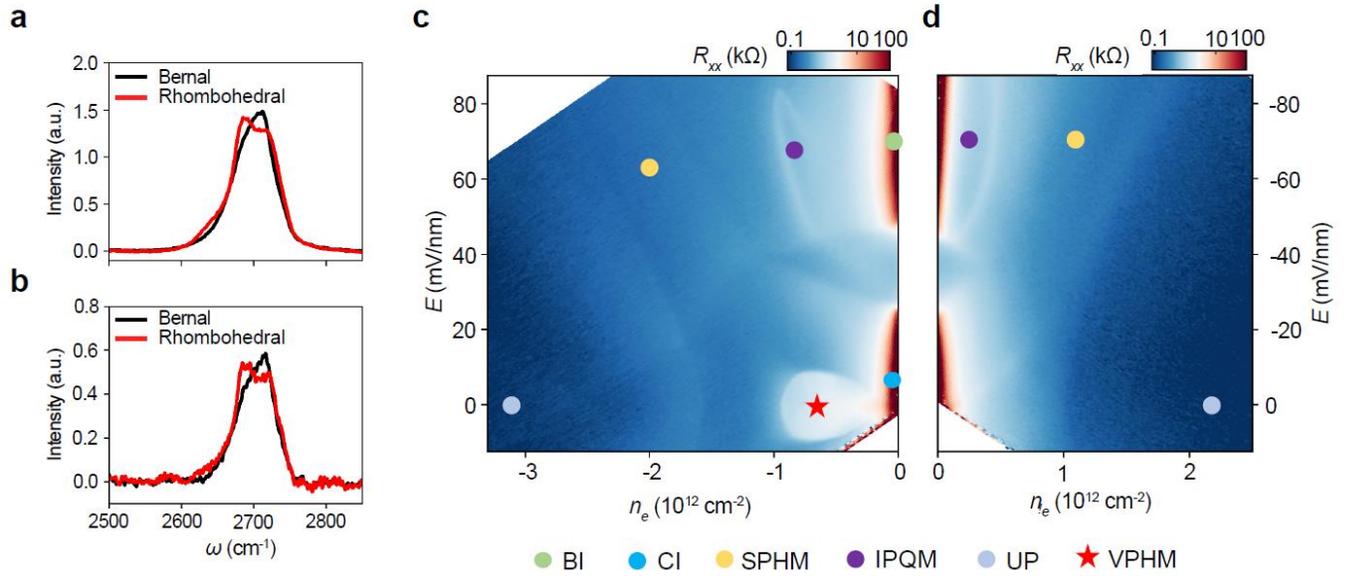

**Extended Data Fig. 1 | Raman characterization and phase diagram of rhombohedral penta layer graphene**

**a, b**, 2D Raman peak of Bernal (black) and rhombohedral (red) stacked pentalayer graphene before (**a**) and after (**b**) hBN encapsulation. The rhombohedral stacking is preserved after hBN encapsulation. **c, d,** Color plots of four-probe resistance $R_{xx}$ as a function of carrier density $n$ and displacement field $E$ for the hole doping side (**c**) and electron doping side (**d**) measured at $B = 0$ and a temperature of 100 mK. Colored dots label different phases including band insulator (BI), correlated insulator (CI), spin-polarized half metal (SPHM), isospin-polarized quarter metal (IPQM) and unpolarized metal (UP). The red star labels the valley polarized half metal (VPHM), which is the focus of the maintext. The details of IPQM and SPHM states can be found in ref[42].

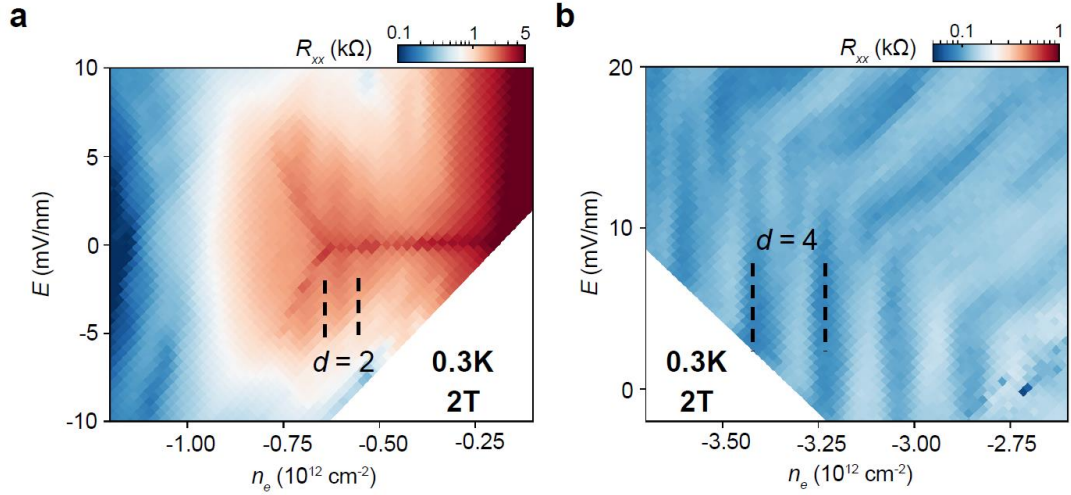

**Extended Data Fig. 2 | Isospin degeneracy inferred from Landau level spacing**

Color plots of four-probe resistance $R_{xx}$ as a function of carrier density $n$ and displacement field $E$ for the small hole doping VPHM region (**a**) and large hole doping UP region (**b**) measured at $B = 2T$ and a temperature of 300 mK. The straight features correspond to Landau levels and the spacing indicates the degeneracy $d$ of the band. With $d = 4$, all four isospin flavors are present at the Fermi surfaces. With $d = 2$, two out of four isospin flavors have Fermi surfaces (the so-called half-metal).

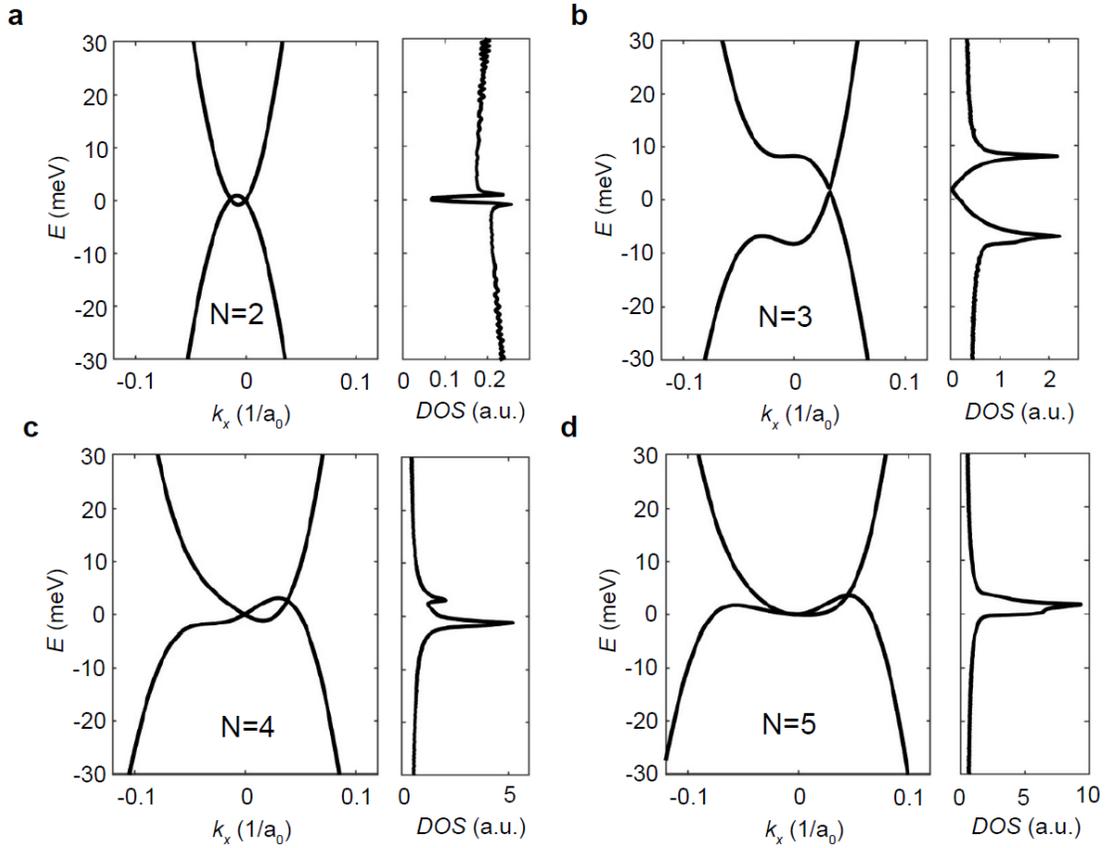

**Extended Data Fig. 3 | Single particle band structure and density of states of rhombohedral stacked multilayer graphene**

Tight-binding calculation of single particle band structure and density of states (DOS) for rhombohedral stacked multilayer graphene (layer number N=2, 3, 4, 5). Due to the remote hopping, the band structure deviates from $E \sim k^N$ at low energy. The rhombohedral pentalayer graphene has the flattest band among all layer numbers.

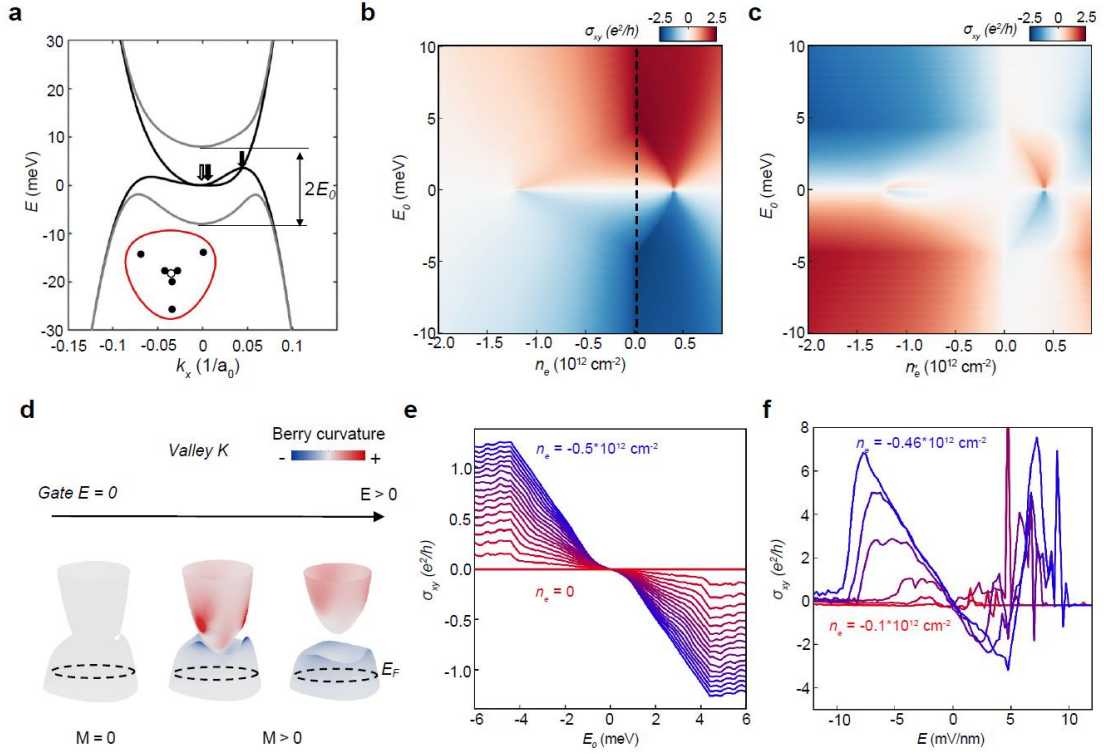

**Extended Data Fig. 4 | Hall conductivity calculation**

**a**, Tight-binding calculation of band structure for rhombohedral pentalayer graphene near the K point with an interlayer potential $2E_0 = 0$ meV (black) and 16 meV (grey). Inset shows an iso-energy contour at –5 meV and the black and white circles represent Dirac points with Berry phase $\pi$ and $-\pi$, which are also labeled by the black and white arrows in the band structure. **b**, Calculation of Hall conductivity $\sigma^{xy}$ with respect to the doping density $n_e$ and interlayer potential $E_0$ for a single valley. The hot spots at around -1.2 and $0.4*10^{12}$ cm$^{-2}$ correspond to the four Dirac points at smaller $k$ and the three Dirac points at larger $k$ respectively. **c**, The same plot as **b** with the contribution from the other valley at zero density (same position and opposite value to the dashed line in **b**). **d**, Colored-coded band structure in the K valley, showing the Berry curvature distribution at varied gate electric field $E$s. The iso-energy contour at Fermi level $E_F$ is labeled by the dashed circle. $\sigma^{xy}$ can be calculated by integrating the Berry curvature below $E_F$. At $E = 0$ mV/nm, Berry curvature is zero everywhere except for at the Dirac points, and $\sigma^{xy}$ is zero. As $E$ increases, Berry curvature hot spots emerge near the Dirac points and spread out. Consequently, there is more Berry curvature on states below $E_F$ and $\sigma^{xy}$ increases with $E$. **e**, Linecuts in c from $n_e=0$ to -$0.5*10^{12}$ cm$^{-2}$ for a single spin copy. At a small interlayer potential $E_0$, the Hall conductivity is roughly linear with $E_0$. **f**, Measured $\sigma^{xy}$ at 20mT and 0.3K. The linear $E$ dependence of $\sigma^{xy}$ agrees with the calculation qualitatively. Measurements on the positive $E$ side suffer from contact issue at low temperatures.

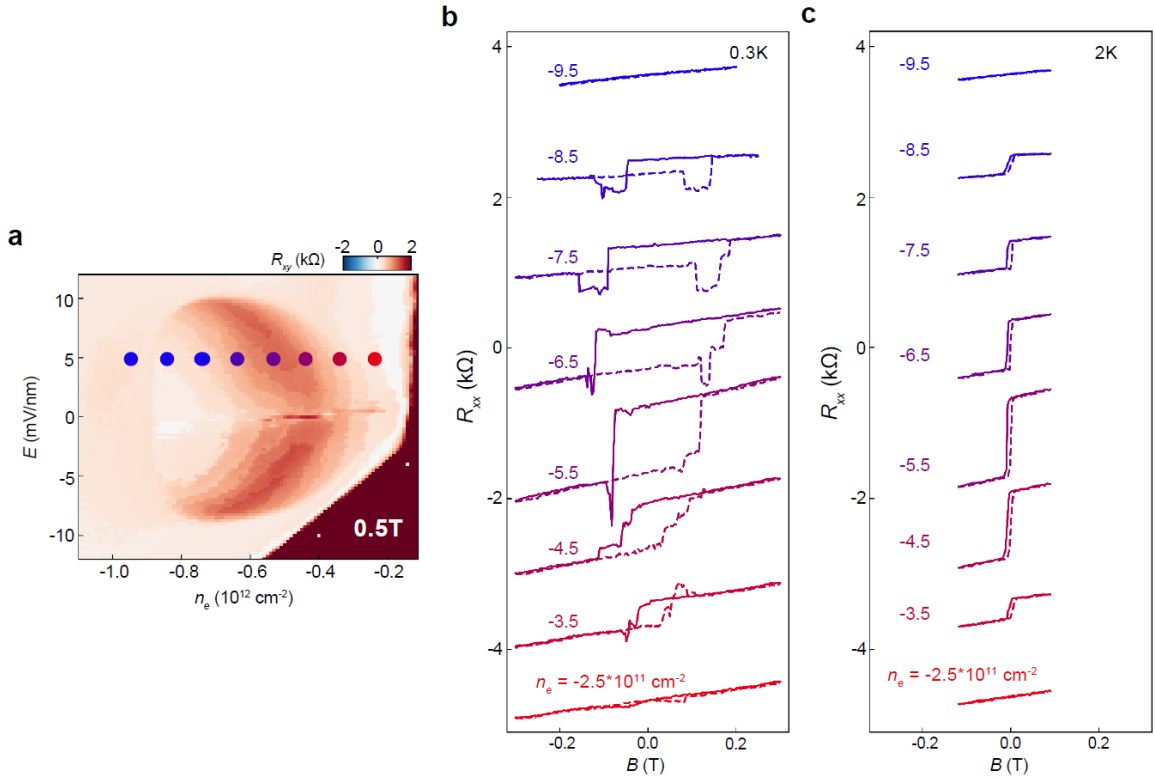

**Extended Data Fig. 5 | Magnetic hysteresis at 0.3K and 2K at different doping**

**a,** Color plots of four-probe resistance $R_{xy}$ as a function of carrier density $n$ and displacement field $E$ for the small hole doping VPHM region at $B = 0.5$ T and $T = 0.3$ K. **b & c,** $R_{xy}$ as the magnetic field $B$ is scanned, showing clear hysteresis loops within the droplet region. Curves correspond to dots with the same color in **a** at $E = 5$ mV/nm and different $n_e$ at $T = 0.3$ K **a** and 2K **b**, and are shifted vertically for clarity. Solid and dashed curves correspond to forward and backward scanning of the $B$ field. At 2K, the coercive field of magnetic hysteresis almost vanishes, consistent with the disappearance of the ferro-valleytronic order at 2K.

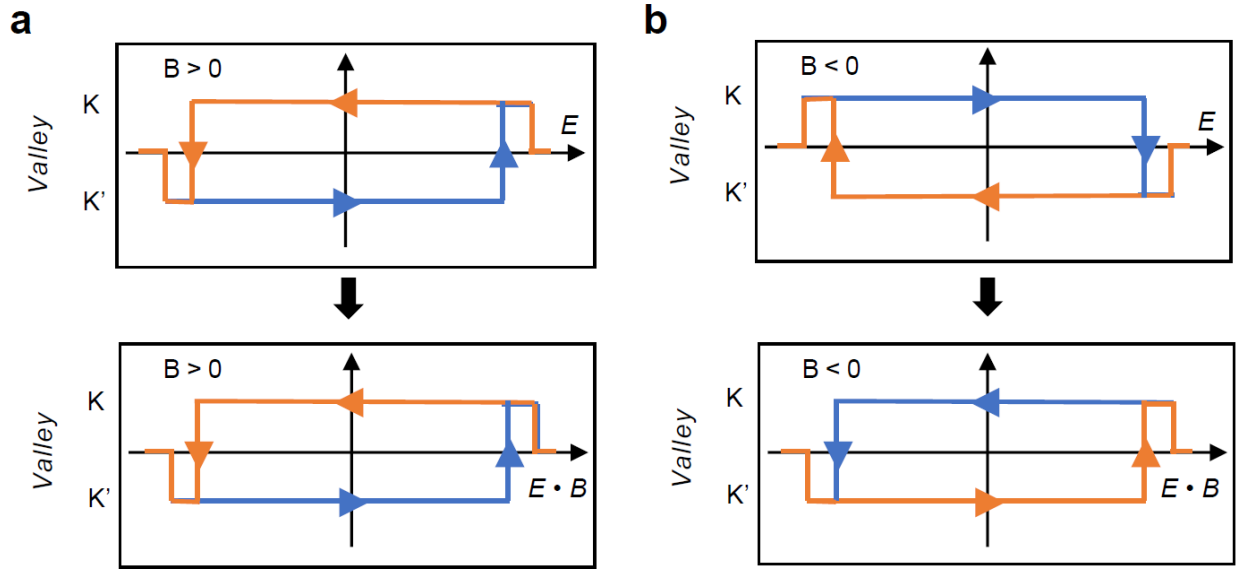

**Extended Data Fig. 6 | Control valley with $E \cdot B$ field**

**a & b,** Plots of valley polarization correspond to the $R_{xy}$ plot in Fig. **3d** and **3e** with $B > 0$ (**a**) and $B < 0$ (**b**), featuring two hysteresis loops winding in opposite ways. However, if we plot the x-axis as the $E \cdot B$ field, the two loops will collapse into one, which demonstrates that the conjugate field of the valley is $E \cdot B$. Blue (yellow) lines labels scanning to positive (negative) $E$ direction. The arrows indicate the scanning direction.

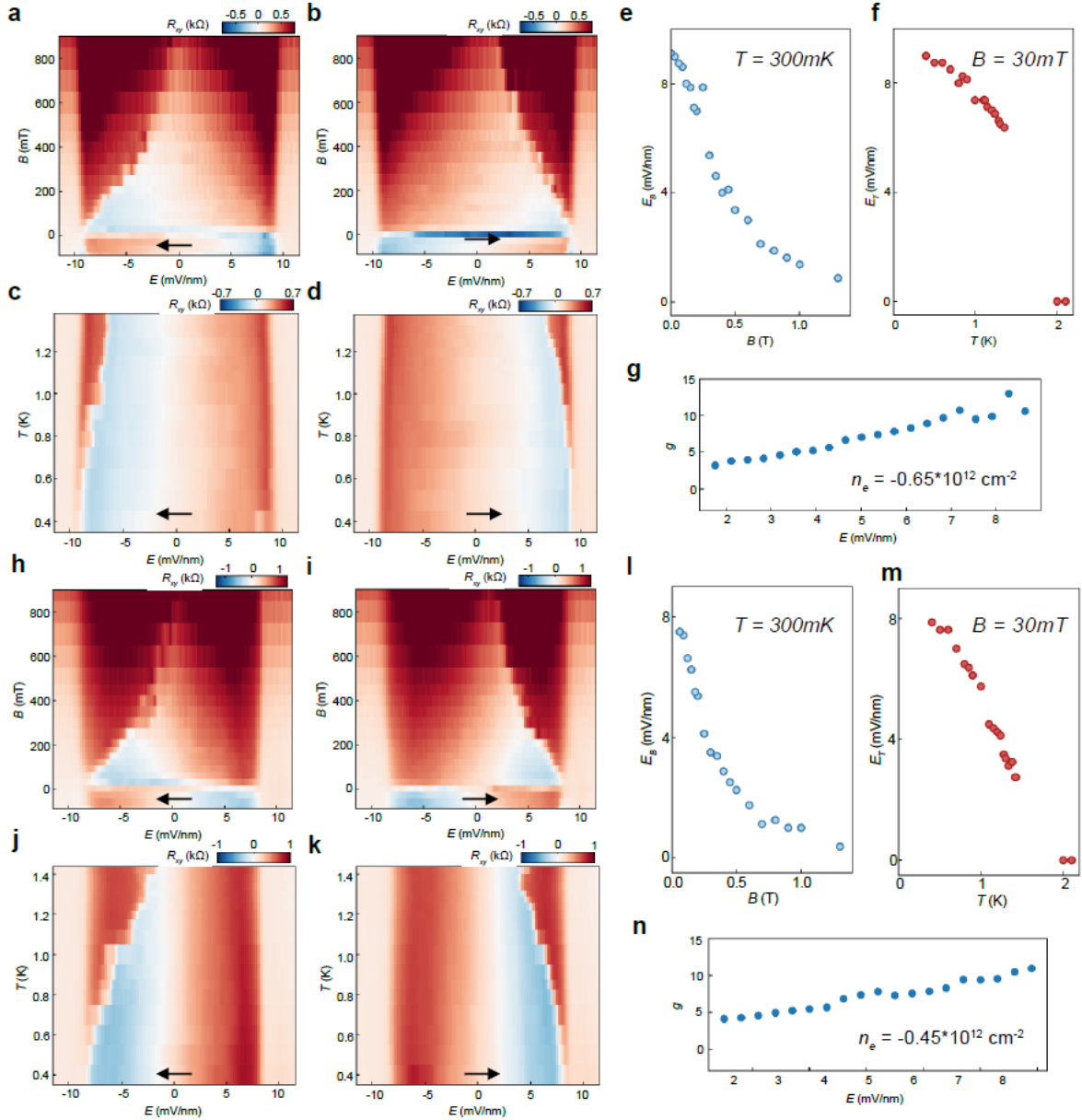

**Extended Data Fig. 7 | Evolution of the 'butterfly' as a function of magnetic field and temperature at $n_e = -0.65*10^{12}$ cm$^{-2}$ and $n_e = -0.45*10^{12}$ cm$^{-2}$. a & b.** $R_{xy}$ as $E$ is scanned at different magnetic fields at $n_e = -0.65*10^{12}$ cm$^{-2}$. **d & e.** $R_{xy}$ as $E$ is scanned at different temperatures at $n_e = -0.65*10^{12}$ cm$^{-2}$ and $B = 30$ mT. **e & f.** Critical fields $E_B$ and $E_T$ from **a-d**. **g.** Effective $g$-factor of the averaged orbital magnetic moment as a function of $E$ extracted from **e & f**. **h-n.** Same measurements and analysis as **a-g** at $n_e = -0.45*10^{12}$ cm$^{-2}$.

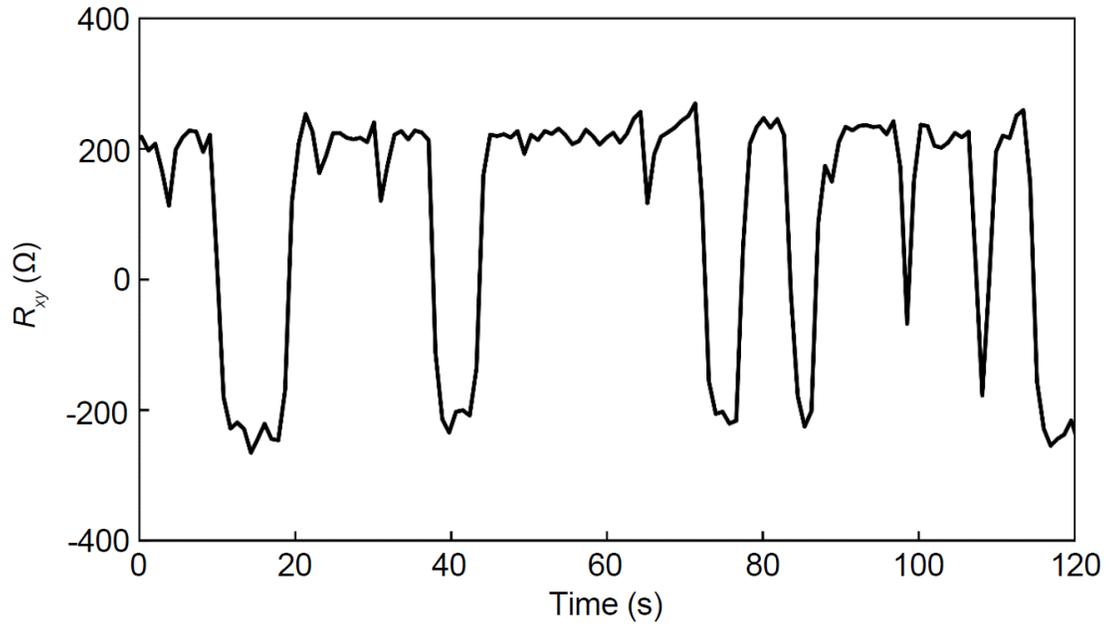

**Extended Data Fig. 8 | Fluctuation of magnetization at zero magnetic field**

$R_{xy}$ as a function of time, measured at $n_e=-0.7*10^{12}\text{cm}^{-2}$, $E = 4$ mV/nm, $B = 0$ and $T = 300$ mK. $R_{xy}$ flips sign frequently with time, indicating the fluctuation of the magnetization due to the finite size of the magnet. A small external $B$ field (20mT) can stabilize the system without changing the ground state.

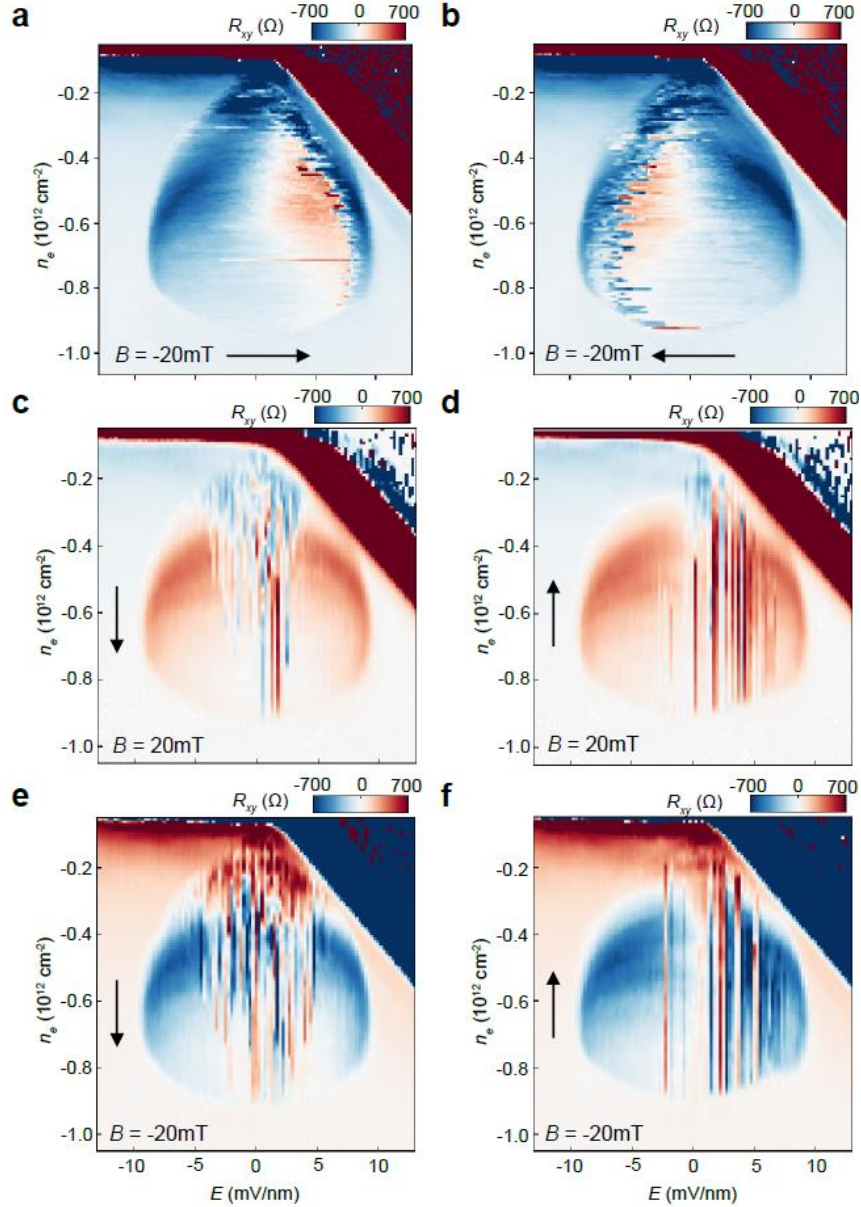

**Extended Data Fig. 9 | Hall resistance behaviors under different measuring conditions**

**a** & **b,** 2D color maps of $R_{xy}$ in the droplet region corresponding to forward and backward scanning of $E$ with a small negative magnetic field $B$ = -20 mT. **c** & **d,** The same plot as in **a** and **b** but with scanning density $n_e$ at $B$ = 20 mT. **e** & **f,** The same plot as in **a** and **b** but with scanning density $n_e$ with $B$ = -20 mT. $R_{xy}$ changes the sign as we scan $E$ but stays the same sign as we scan $n_e$. This is due to the fact that the magnetic moment for a given valley will reverse as we scan $E$ but not $n_e$. As we scan n, the system will choose to polarize to the valley with the magnetic moment parallel with external $B$. The random jumps near $E$ = 0 when scanning $n_e$ is because the magnetic moment near $E$ = 0 is very small and the coupling to the external B field is weak, so the system cannot decide which valley to choose.

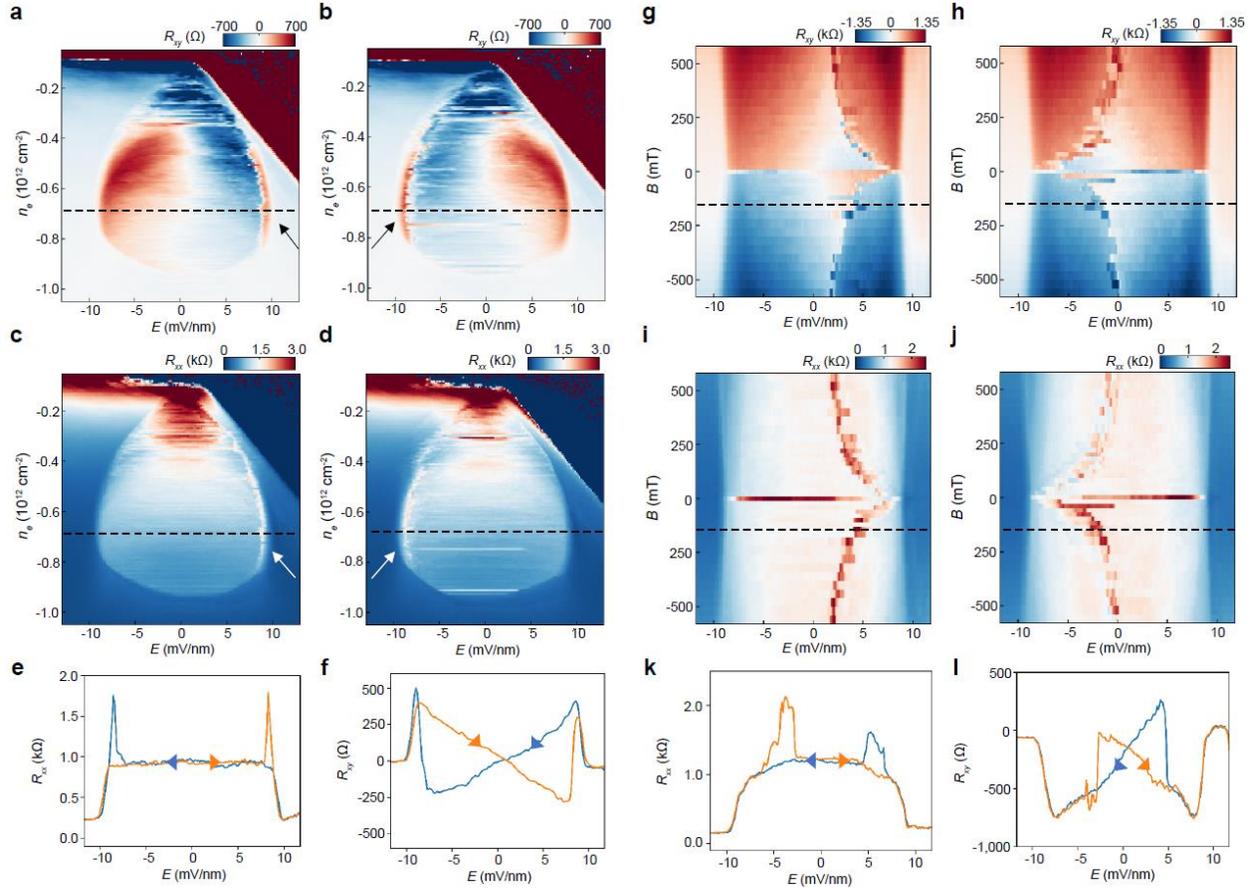

**Extended Data Fig. 10 | Longitudinal resistance $R_{xx}$ hysteresis**

**a** & **b**, The same plot as Fig **3d** & **e**. The arrow indicates the jump in $R_{xy}$ where the valley switches. **c** & **d**, The same plot as in **a** & **b** for $R_{xx}$. The arrow indicates where $R_{xx}$ jumps and it also coincides with where the $R_{xy}$ jumps. **e** & **f**, Linecuts of $R_{xx}$ and $R_{xy}$ at $n_e=-0.67*10^{12}\text{cm}^{-2}$ as indicated by the dashed line in **a-d**. **g** & **h**, The same plot as Fig **4a** & **b**. **i** & **j**, The same plot as in **g** & **h** for $R_{xx}$. The jumps in $R_{xx}$ correspond to that in $R_{xy}$. **k** & **l**, Linecuts of $R_{xx}$ and $R_{xy}$ at $B = -0.16$T as indicated by the dashed line in **a-d**. The jumps in $R_{xx}$ indicate the valley-switching is a first-order process.

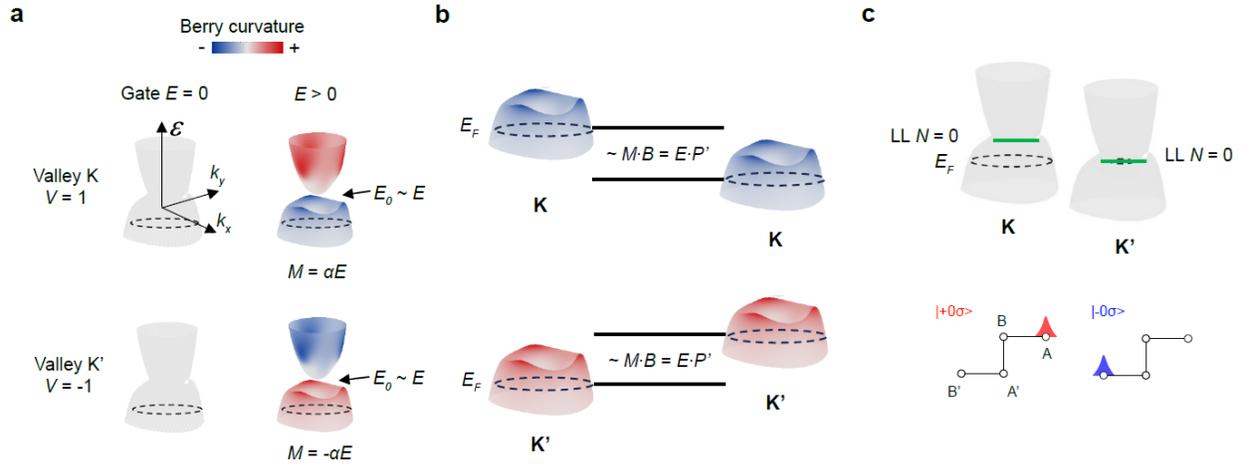

**Extended Data Fig. 11 | Ferro-valleytricity and ferroelectricity**

The opening of the band gap $E_0$ by a gate electric field $E$ is effectively inducing an electric dipole $P_0$ which couples to $E$ and lowers the energy of the occupied states. The Berry curvature and orbital magnetization in the K and K' valleys are opposite. **b.** Given a non-zero $E$, the valence bands in the K and K' valley experience further shift in a magnetic field $B$. The shift is proportional to the orbital magnetization of the occupied states. This shift in energy $M*B = aE*B$ can be viewed as $M*B = E*(aB) = E*P'$, where is effectively an additional electric dipole $P'$. This latter part of electric dipole $P'$ matters for the valleytricity while the $P_0$ does not. **c.** Ferro-electricity in the valley-polarized state at $E = 0$, $|B| > 0$. Upper panel: illustration of the valley polarization and imbalanced population of the zeroth Landau levels in K and K' valley at a non-zero $B$. Lower panel: valley and sublattice and layer have a one-to-one-to-one correspondence in the zeroth Landau level of rhombohedral graphene. Thus a valley polarization at non-zero B implies layer polarization and an electric dipole at $E = 0$ mV/nm. In this picture, ferro-electricity should exist at $E = 0$ mV/nm and $|B| > 0$ mT.

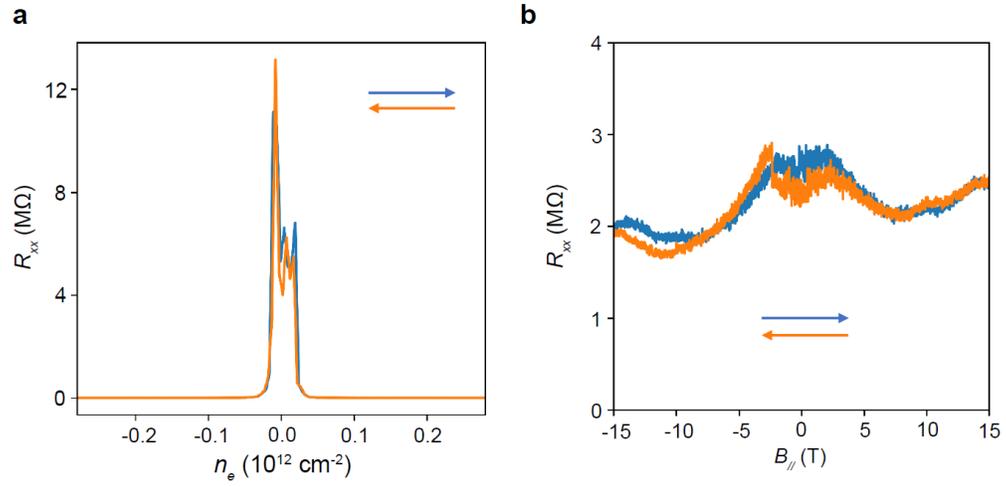

**Extended Data Fig. 12 | Density and in-plane magnetic field scan at $E = 0$**

**a.** $R_{xx}$ versus forward and backward scanning of $n$. **b.** $R_{xx}$ versus forward and backward scanning of $B_{//}$.